# LabVIEW is faster and C is economical interfacing tool for UCT automation

Ankur Kumar[1, a)] and Mayank Goswami[1, b)]

*[1]Divyadrishti Laboratory, Department of Physics, IIT Roorkee, Haridwar, Uttarakhand, India*

[a)] ankur89587@gmail.com
[b)] mayank.goswami@ph.iitr.ac.in

## Abstract

An in-house developed 2D ultrasound computerized Tomography system is fully automated. Performance analysis of instrument and software interfacing soft tools, namely the LabVIEW, MATLAB, C, and Python, is presented. The instrument interfacing algorithms, hardware control algorithms, signal processing, and analysis codes are written using above mentioned soft tool platforms. Total of eight performance indices are used to compare the ease of (a) realtime control of electromechanical assembly, (b) sensors, instruments integration, (c) synchronized data acquisition, and (d) simultaneous raw data processing.

It is found that C utilizes the least processing power and performs a lower number of processes to perform the same task. In runtime analysis (data acquisition and realtime control), LabVIEW performs best, taking 365.69s in comparison to MATLAB (623.83s), Python ( 1505.54s), and C (1252.03s) to complete the experiment. Python performs better in establishing faster interfacing and minimum RAM usage.

LabVIEW is recommended for its fast process execution. C is recommended for the most economical implementation. Python is recommended for complex system automation having a very large number of components involved. This article provides a methodology to select optimal soft tools for instrument automation-related aspects.

## Introduction

The instrumentation and automation have always supported the development of science & research for industrial growth. Generally, the automation requires interfacing of various instruments, sensors, and electromechanical components to work in coordination and perform predefined tasks with fewer manual interventions. Automation provides the convenience of performing various monotonous tasks quickly and repeatedly with relatively better precision and accuracy. With the advancement of microcontrollers, integrated circuits, sensors, and electronics, the development of custom-designed systems for scientific research is increasing[1]. Any automated system requires a set of algorithms to govern its electromechanical components. These set of algorithms activate integrated sensors and actuators in a predefined synchronized fashion. Simultaneously, the output data is acquired by sending certain signals for stimulation and saved in memory. The overall control system also keeps track of all processes via a feedback loop to correct any deviation. The signal processing to extract and comprehend the essential information is performed by another set of algorithms. The set of algorithms, for activation of (a) sensor, (b) actuator, (c) feedback loop along with (d) signal and image processing may or may not be written in the same programming languages / soft tool platform. The choice of these tools generally depends upon the developer's level of comfort and may not be optimal as far as the performance of the overall design is concerned. While developing a fully automated ultrasound computed tomography (UCT) system, we ourselves faced the dilemma if we would be using the optimal soft tools, otherwise costing us the overall performance.

The soft tools comparison for the instrument automation is least explored. In literature, the soft tools are compared for different fields of applications. In [2], analysis is performed to conduct an empirical study to analyze the productivity variations across different programming languages. In bioinformatics, three programming languages for a full-fledged next-generation sequencing tool are compared [3]. The memory usage and speed of execution for three standard bioinformatics methods for six different programming languages are compared [4]. In macroeconomics, several programming languages are compared to solve stochastic models [5]. The existing studies are primarily focused on the software point of view without including the automation of any robotic system.

The criteria for selection of soft tools may include a convenient user interface, availability of a wider user base(online forums), richness in an already developed function library, runtime performance, compactness in the length of codes, RAM, memory, and processor utilization[6, 7]. The performance of the programming language's IDEs has been studied in the various aspects of instrument interfacing, data acquisition, instrument control, and data processing [6, 8].

To choose the best option for our UCT automation, we have compared the performance of two commercially available programming platforms, namely: LabVIEW™, MATLAB®, and two computer programming languages: Python and C [6, 9-11]. The comparison is carried out by developing the codes to perform specific tasks individually and collectively. The performance parameters are (a) runtime efficiency, (b) performing the number of input/output processes, (c) processor utilization, and (d) RAM usage during the data acquisition, reading/writing data, data processing, and realtime control. These performance indices are expected to quantify the resource utilization by any given soft tool. This study focuses on comparing the mentioned programming soft tools to investigate the efficiency of performing different tasks.

### Motivation:

Instrumentation of a system includes various aspects such as components interfacing, data measurement, processing, electromechanical controls, etc. Generally, research groups just go for the programming languages they are familiar with, resulting in limited efficiency and sub-optimal performance. It may also affect the quality of the output results (sometimes, we overlook the precision). The articles that focus on the performance evaluation of multiple automation soft tools for a given system together are rare. To the best of our knowledge, we could not find an article in the literature that evaluate the interfacing soft tools' performances while testing the respective codes in realtime with instrumentation development. This article provides a methodology to analyze the performance of soft tools.

## Methodology

### Brief details about the UCT system

The automated UCT system comprises two non-contact ultrasound (NCU) transducers, an arbitrary wave generator (AWG), a digital storage oscilloscope (DSO), a microcontroller, electromechanical components, and a processing system, as shown in figure 1. The AWG, DSO, and microcontroller are connected via a USB port. The AWG is coupled to one ultrasound transducer, while DSO is coupled to another via BNC cables. A cheap but reliable microcontroller Arduino UNO R3 may be used to control actuators in realtime. The performance of soft tools, however, is independent of the choice of any microcontroller. All four soft tools' performance is evaluated using the same microcontroller. The microcontroller is connected to the actuators via driver electronics based on an H-bridge circuit to control the motor's shaft precisely. It also provides the convenience of setting micro-stepping resolution as per the requirement. The electromechanical components consist of the rotating table and the linear translation platform coupled with linear ball bearings and a threaded-rod system.

The transducers are placed into their respective stationary 3D printed holders on this platform. The object (to be scanned) is placed on the rotating and sliding table. In our design, AWG generates an input pulse to trigger one ultrasound transducer (called emitter henceforth). In turn, the transducer produces ultrasound waves that traverse through the object and are detected by the other transducer (termed as receiver) placed at the other end of the object. The DSO, coupled with the receiver, acquires the raw data and performs necessary analog to digital conversion with a very high sampling frequency of $10 \times 10^9$ Samples/s (GS/s). The parallel beam geometry is applied to acquire and process the scanning data. The process is repeated multiple times while linearly translating and rotating either the transducer pair or object to reduce the data sparsity. The acquired data is processed to reconstruct the specimen profile [12]. The same algorithms are used to keep the analysis consistent for all soft tools.

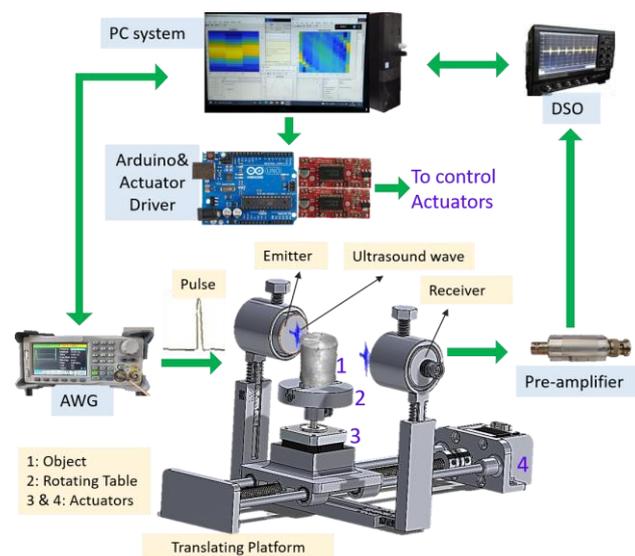

*Figure 1: The automated UCT system.*

The automation codes are executed on the same PC. The Dell Inspiron Desktop computer equipped with i5 6400 microprocessor and 8 GB RAM is used. The system is installed with the Windows 10 (21H2) 64-bit operating system.

Instrument Interfacing:- The above-mentioned instruments can be connected to the PC through USB Test and Measurement Class (USBTMC), Serial Communication, Wi-Fi, Ethernet, IEEE 488.2I(GPIB), etc. VISA (Virtual Instrument Software Architecture), an instrument driver, is installed on the PC to facilitate communication. VISA is a standard for configuring, programming, and troubleshooting instruments comprising GPIB, VXI, PXI, serial (RS232/RS485), Ethernet/LXI, and/or USB interfaces [13, 14]. It includes utilities and low-level control features required to control the instruments. The AWG and DSO are connected to the system via the USBTMC interface. In DSO, the connection via USBTMC is configured by selecting the utilities >setup>Remote>USBTMC in the DSO. The AWG has a plug-and-play interface and does not require any setting selection. The microcontroller is controlled through a UART (Universal Archonous Receiver Transmitter) serial communication interface via a USB connection[15, 16].

The instruments can be interfaced with PC using soft tools either by developing custom algorithms or by using the instrument drivers provided by the manufacturers[17, 18]. These drivers offer convenience in controlling and automating the instrument by translating the algorithms into Standard Commands for Programmable Instrumentation (SCPI) commands that the instrument can understand[19, 20]. These drivers provide the utility for the conversion of the received raw data into the numeric form. In our case, these drivers are made available for the MATLAB and LabVIEW only by AWG and DSO manufacturer support websites. So, custom algorithms for input/output operations using SCPI commands are developed for automation for C and Python IDEs. These custom algorithms are developed using instrument commands for input-output operations and implementing the data conversion algorithms. The availability of these drivers and interfacing types used to establish the connection for the respective languages is tabulated in Table 1. The Arduino is controlled by a serial communication interface. Whenever the coded instructions are sent from the soft tool to the Arduino, it decodes them into a set of instructions, then guides the Arduino to send the signal accordingly. Arduino IDE (1.8.19) is used to write and deploy the serial interfacing code. It is ensured that all IDEs have no cross dependencies.

*Table 1: Interface type and Driver availability.*

| Soft-Tool (language/Platform) | IDE | Instrument | Interface type (USB Based) | Driver/Firmware Availability |
|---|---|---|---|---|
| C | Visual Studio(VS) v17.00 *Compiler*: gcc 10.3.0 (Rev5, Built by MSYS2 project) | AWG | USBTMC | Custom designed |
| | | DSO | USBTMC | Custom designed |
| | | Arduino | Serial Communication | Custom designed |
| Python | Python 3.9.6 (Spyder) | AWG | USBTMC | Yes |
| | | DSO | USBTMC | Custom designed |
| | | Arduino | Serial Communication | Custom designed |
| MATLAB | MATLAB R2019b | AWG | USBTMC | Yes |
| | | DSO | USBTMC | Lecroy_basic_hr_driver |
| | | Arduino | Serial Communication | Custom designed |
| LabVIEW | LabVIEW 2018 | AWG | USBTMC | SDGX LabVIEW Driver 1.0.1 |
| | | DSO | USBTMC | IVI Driver 3.2.9.0 x64, LeCroy VICP Passport |
| | | Arduino | Serial Communication | LIFA_base |

Data measurement:- The 12-bit raw data is acquired while scanning the object by varying the number of detectors and projections. The acquired data is written to a .txt file. The size of these files depends on the sampling rate, the detectors and/or projections. The acquired data file contains the time information and the amplitude data of the signal. The study is carried out to analyze the performance in acquiring and reading/writing the acquired data with variations in data size.

Data processing:- Data processing codes are developed in each programming language using the same algorithms. Data processing mainly involves the algorithms to process the ultrasound signal to extract the meaningful data called projection data. Projection data is then used to reconstruct the object's profile via a separate CT algorithm.

The programming codes for all languages are written in the same approach from scratch to minimize the difference. The main code for the analysis is written in two forms: a) without signal processing and b) with signal processing. Multiple experiments using the developed UCT system are performed to assess the performance of soft tools. The process flow diagram is shown in figure 2. The experiment is carried out to scan the object for 40 rotations and 40 linear translations. Pulse waves of the duty cycle of 4.35%, amplitude of 20V, and pulse width of 2.9e-08 s at a frequency of 1.5 MHz are used to trigger the emitter. In turn, the emitter produces low-power ultrasound waves to scan the object. Low power ultrasound waves having intensity up to $100 \, mW/cm^2$ are used for NDT applications as these waves are elastic in nature and cause no harm to the propagating medium[21]. For each rotational angle, 40 linear translations are performed, and data is acquired in each, resulting in a total of 1600 files. Initially, communication to all system modules is established then the data is acquired for each linear translation movement for all the rotation angles. The acquired raw data is saved to the specified directory for further analysis. When the data processing is enabled, acquired data is processed onboard to extract meaningful data after each acquisition. This extracted data is used to generate the partial tomograph after each rotation. The performed analysis is visualized on the screen after each rotation. When all the data is processed, final images are projected on the screen and saved in the system. The timing information for each major part of the code is recorded. A video collage (Video) is provided to showcase the full scanning process by all four soft tools.

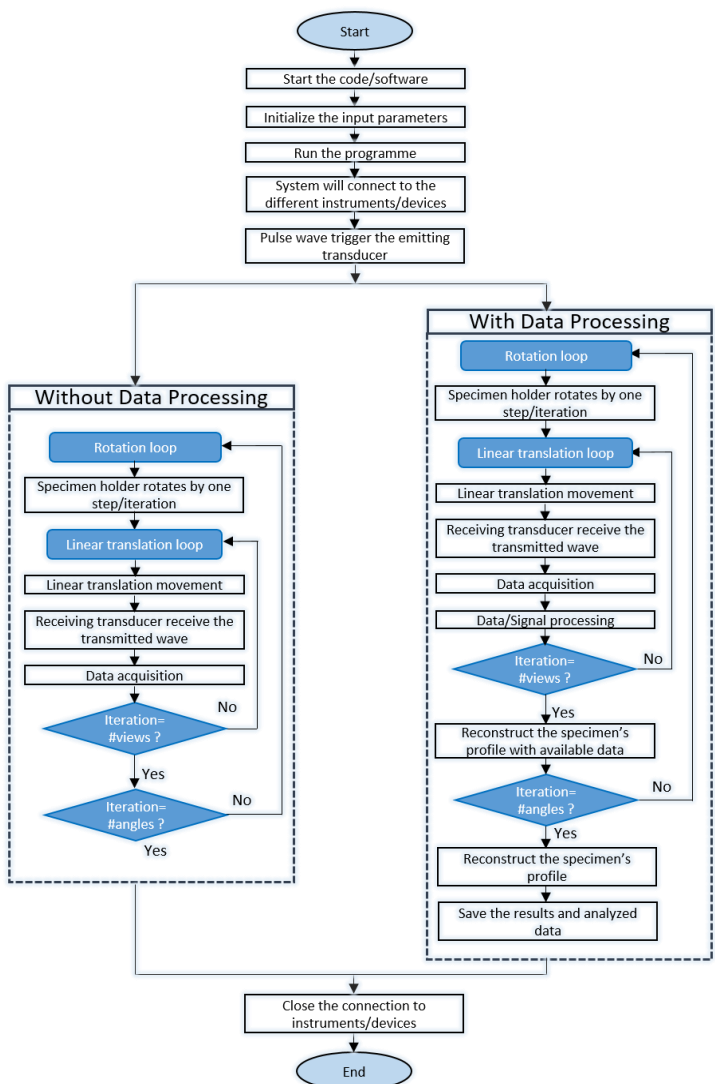

*Figure 2: Process Flow Diagram*

## Performance Indices:

The performance of the soft tools is studied by following parameters:-

1. **IDE launch performance:** Each IDE is started after a fresh restart of the system, then it is started several times. The time was recorded until the IDE became responsive. It includes the time taken by the IDE to load background processes for proper functioning. The launch time represents the time taken until the IDE main window appears on the screen, while the response time represents the time until the IDE starts responding to the user commands. It adds up to the overall user experience.

2. **Performance analysis:** Processor utilization(% Processor Time), RAM(Private bytes) usage, IO(Input/Output) Data Bytes/sec, IO Data Operations/sec, IO Read Bytes/sec, IO Read Operations/sec, IO Write Bytes/sec, IO Write Operations/sec are the parameters used to analyze the performance [22]. All these eight parameters are recorded while the experiment is running on the system. The codes are executed for both forms with and without signal processing. A performance monitoring tool integrated into the PC's operating system is used to estimate these parameters[23].

Brief details about the selected parameters: -

i.   *Processor utilization:* Processor Time is a measure of processor utilization by a specific application. It is the percentage of elapsed time that all of the process threads used the processor to execute instructions. Code executed to handle some hardware interrupts and trap conditions are included in this count.

ii.  *RAM usage:* Private Bytes is the current size, in bytes, of memory that an application has allocated that cannot be shared with other applications.

iii. *IO Data Bytes/sec*: The rate at which the process reads and writes bytes in I/O operations. This counter counts all I/O activity generated by the process to include file, network and device I/Os.

iv.  *IO Data Operations/sec*: The rate at which the process is issuing read and write I/O operations. This counter counts all I/O activity generated by the process to include file, network and device I/Os.

v.   *IO Read Bytes/sec:* The rate at which the process is reading bytes from I/O operations. This counter counts all I/O activity generated by the process to include file, network and device I/Os.

vi.  *IO Read Operations/sec:* The rate at which the process is issuing read I/O operations. This counter counts all I/O activity generated by the process to include file, network and device I/Os.

vii. *IO Write Bytes/sec:* The rate at which the process is writing bytes to I/O operations. This counter counts all I/O activity generated by the process to include file, network and device I/Os.

viii. *IO Write Operations/sec:* The rate at which the process is issuing write I/O operations. This counter counts all I/O activity generated by the process to include file, network and device I/Os.

**Runtime Performance:-** The runtime performance of all the soft tools is recorded. The experimental codes are divided into multiple sections: i) for establishing the connection to the instruments: AWG, DSO and Arduino, ii) for measurements for a single rotation with and without signal processing (40 acquisitions in a single rotation, total of 1600), iii) for generating graphics of the processed data in each rotation and iv) total time for the complete experiment. In another experiment, data acquisition, read and write timing performance is recorded for 100 data sets with increasing data contents. It helped in analyzing the performance and consistency of the soft tools with an increase in data loads. For reading/writing data, the basic .txt format is used for efficiency.

## Results and analysis

The performance of the soft tools is analyzed in several aspects of the automated scientific device. It includes interfacing with the instruments/devices, data acquisition, and data processing. Data processing includes generating the graphics, representing the processed data, and several read/write operations. An in-house developed UCT system is used.

### 1. IDE launch performance:

The average launch time is estimated in two steps: (with and without restarting the PC, termed as first launch time (T1) and second launch time (T2), respectively. The exercise to record T1 and T2 is executed six times for each IDE. The mean values are plotted in figures 3 a) and 3 b). The standard deviation values were found to be less than 0.8 s for all the soft tools. In the first launch (T1), the IDE for C language, i.e., VS, took the least launch time, while the Spyder IDE for Python took the maximum time. When these IDEs are started second time (T2) without restarting the PC, VS again showed a faster launch time while LabVIEW showed a faster response time.

In every fresh start(T1), each of the IDE took significantly more time than they required when started again(T2), suggesting that several background processes are running (associated with IDE) even after closing it. It may have enhanced the launch performance of the respective IDEs. This behavior is consistently observed for all IDEs.

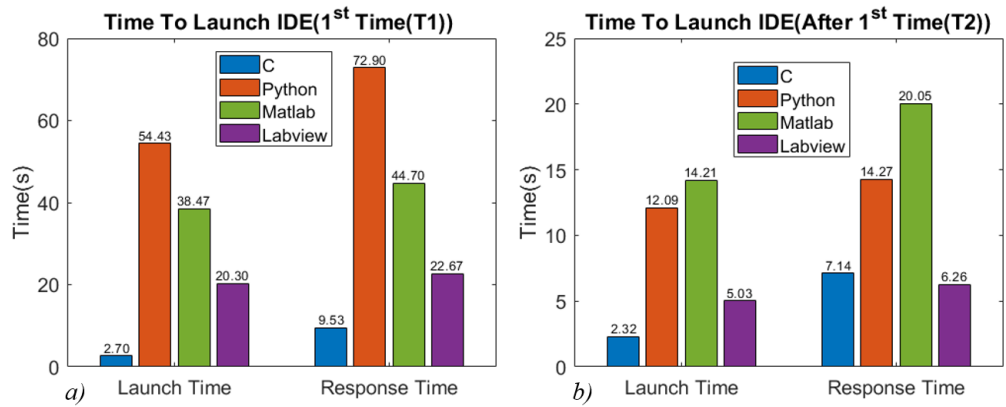

*Figure 3:  IDE launch performance a) average time to Launch IDE in fresh start(1st time) (T1) and average time until IDE became responsive, b) Average time to Launch IDE after 1st time (T2) and average time until IDE became responsive.*

## 2. Performance analysis:

The plots for the performance analysis of soft tools show the parameter variation during the experiment. The x-axis represents the time taken by the IDE to complete the process. The y-axis represents the respective process. The vertical bar shows the standard deviation ($\sigma$), and the black dot shows the mean ($\mu$) value of the process.

The processor and RAM usage with and without data processing are shown in figure 4. The C consumes the least processing power to perform the same analysis while MATLAB consumes the most, as shown in figure 4 a). However, in performing the same processes, Python acquired the least RAM while MATLAB acquired the most, as shown in figure 4c). Also, processor and RAM usage increased slightly for all languages when the data was also processed during the experiment. The standard deviation was least for C while highest for MATLAB for both processor utilization and RAM usage. In the case of MATLAB, the RAM usage increased as the experiment advanced, suggesting that more memory was acquired to store the working data and the machine codes.

The IO processes give in-depth information about the IO processes executed during the experiment. The IO read/write bytes/sec and operations/sec are shown in figure 5. The minimum number of processes/sec are executed for C during the experiment for all the indices in both of the cases (With and without data processing). In the experiment without data processing, LabVIEW executed the higher number of IO read operations/s, IO read bytes/s, IO write bytes/s, IO data bytes/s and IO data operations/s (as shown in figures *5a), e), g)* and figure

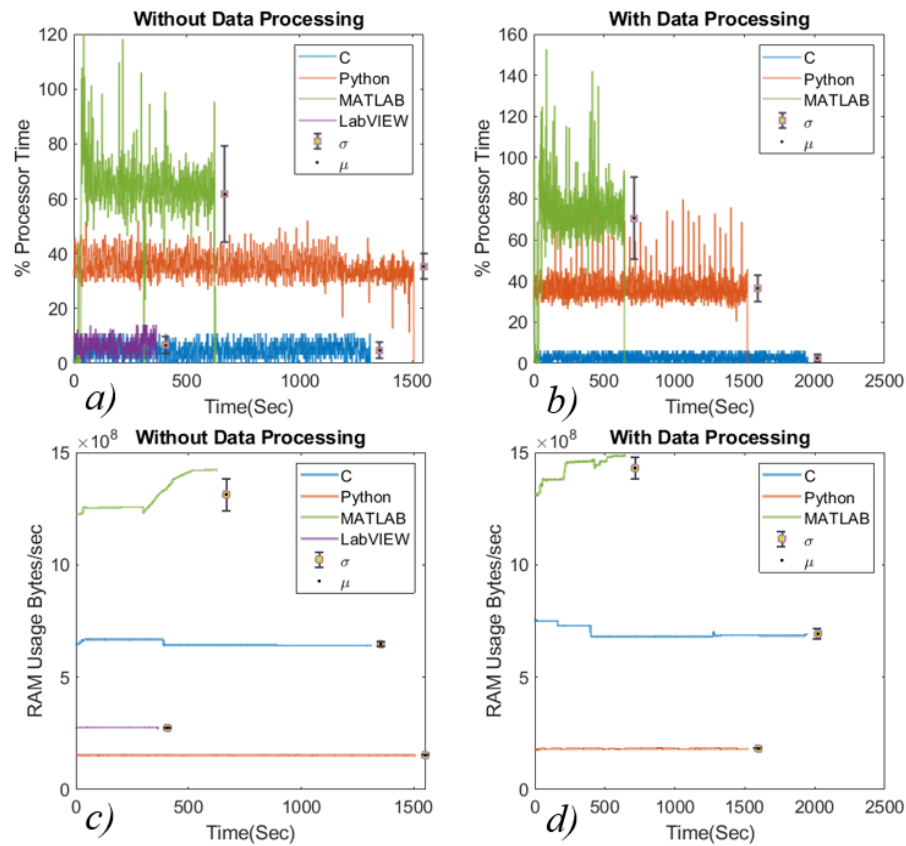

*Figure 4: Performance analysis based on a) & b)Processor utilization and c) & d) RAM usage during the experiment.*

*6a),c),* respectively) and MATLAB has executed a higher number of IO write operations (figure *5c)*). However, when the data is processed, Python executed a higher number of IO read bytes/s, IO read operations/s and IO data operations/s (refer to figures *5b)*, *f)* and *6d)*) while MATLAB executed a higher number of IO read bytes/s, IO read operations/s and IO data operations/s as shown in figures *5d)*, *h)* and *6b)*. The mean values of the performance analysis parameters are tabulated in table 2. It highlights the best values of performance indices values.

The analysis shows that the C performed the least number of IO processes/s, which is also supported by the least processor utilization during both the experiments (with and without data processing). LabVIEW executed a higher number of processes/s in 5/6 IO processes during the experiment without data processing. MATLAB and Python have executed a higher number of processes/s in three IO processes each during the experiment with data processing. As the experiment ends, the number of IO processes along with the processor utilization reduces to a minimum value. However, the RAM usage didn't decrease even though the processes were reduced to a minimum, suggesting that the IDEs reserve RAM.

The standard deviation was the least for the C language. Python has shown a higher standard deviation in the experiment with data processing. The mean value of processes for all the soft tools didn't significantly change for both the cases, i.e., with and without data processing. The time taken by the soft tools was more when the experiment was carried out with data processing, as evident from the IO process plots and shown in figure 9 *a)* & *b)*. In Python, the processor utilization and IO processes/s bounced several times, matching the number of times the IDE is rendering the graphics. It suggests that Python executes more background processes/s to generate the graphics.

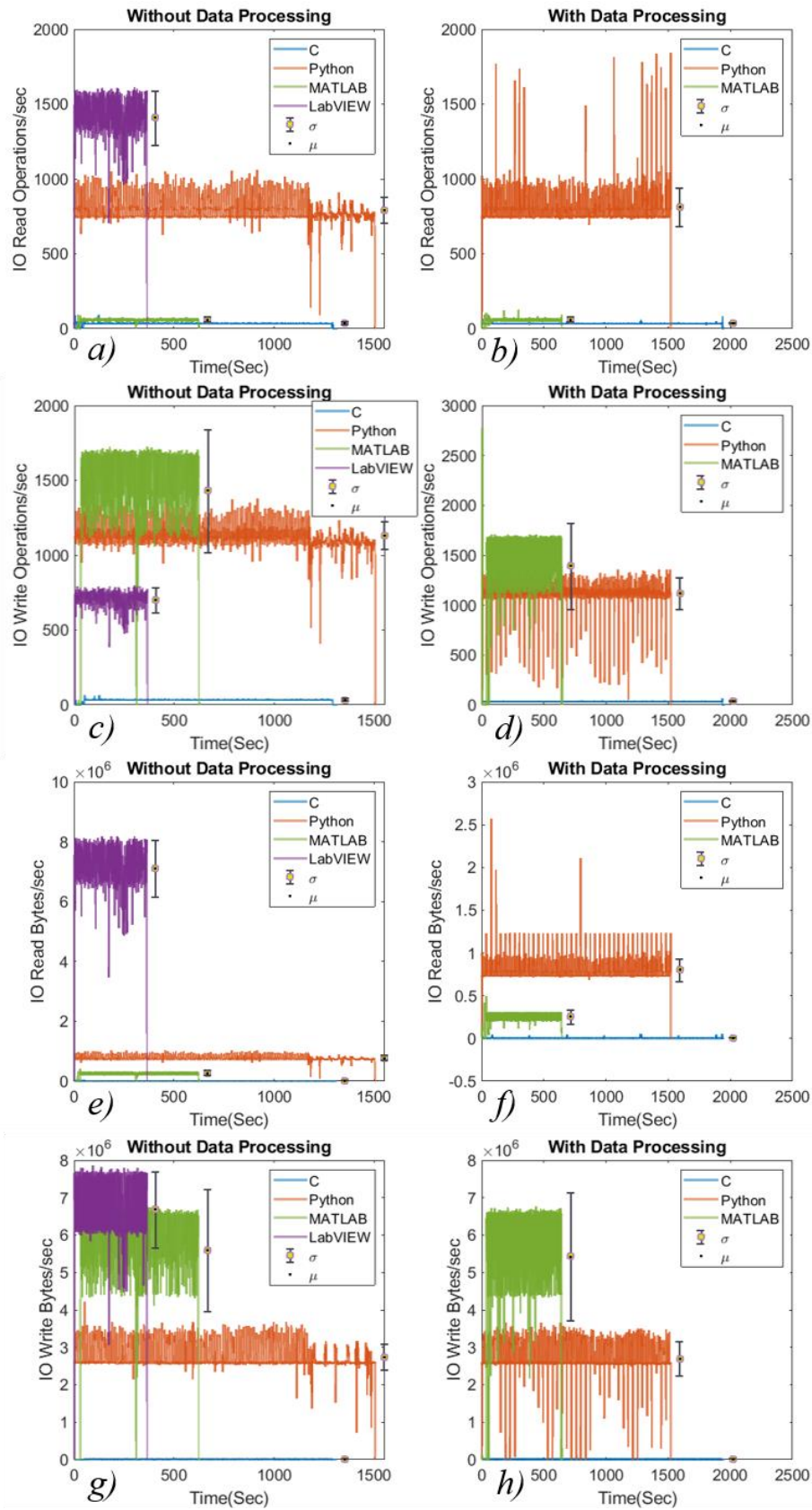

*Figure 5: IO read/write bytes and operations executed during the experiment, vertical bar represent the standard deviation(σ) and black dot inside the circle represent the mean(μ) value .*

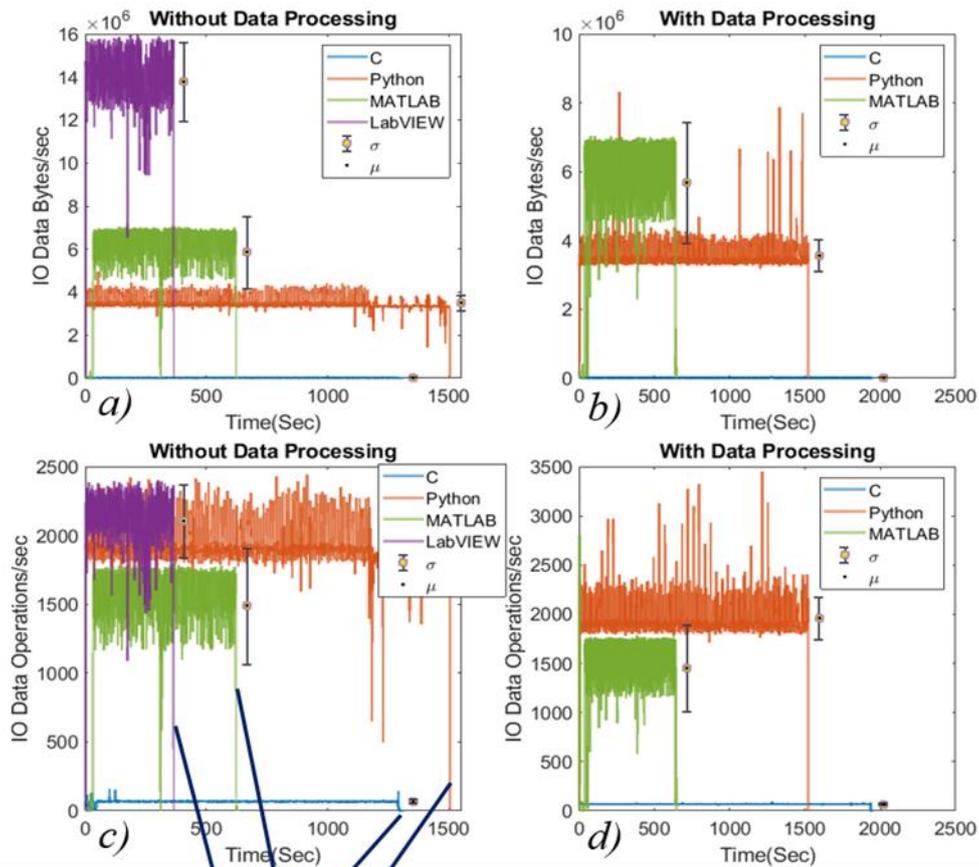

Figure 6: IO data bytes and operations during the experiment.

Table 2: Mean value of parameters used for performance analysis

| IDE | Data Processing | Processor Util. | RAM Usage Bytes/s x10^8 | IO Processes | | | | | | |
| | | | | IO Data Bytes/s x10^4 | IO Data Operations/s | IO Read Bytes/s x10^4 | IO Read Operations/s | IO Write Bytes/s x10^2 | IO Write Operations/s |
| C | Enabled | 3.28 | 6.94 | 4 | 75.66 | 3.09 | 41.55 | 90.71 | 34.11 |
| | Disabled | 5.54 | 6.49 | 3.14 | 76 | 2.25 | 42.44 | 89 | 33.55 |
| Python | Enabled | 36.74 | 1.81 | 393.31 | 1958.6 | 123.01 | 839.81 | 27029.76 | 1118.78 |
| | Disabled | 35.3 | 1.53 | 351.98 | 1925.32 | 77.53 | 792.26 | 27444.88 | 1133.05 |
| MATLAB | Enabled | 72.07 | 14.32 | 568.31 | 1455.35 | 25.29 | 59.13 | 54301.95 | 1396.22 |
| | Disabled | 63.2 | 13.13 | 586.39 | 1488.83 | 25.79 | 57.68 | 56060.53 | 1431.15 |
| LabVIEW | Disabled | 6.929 | 2.76 | 1379.87 | 2107.92 | 711.66 | 1407.99 | 66820.27 | 699.93 |

## 3. Runtime Performance

1. Data acquisition, reading and writing speed are measured for 100 data sets with the increase in data contents as shown in figures 8*a), b),* and *c)*. Data sets containing 1k, 2k, 5k, 10k, 20k, and 50k data points are acquired through the DSO, and the same data sets are read and written to .txt files. In data acquisition, the soft tool's performance is mixed; C performed better in acquiring data sets with contents up to 5k, while LabVIEW performed better for 10k and 20k datasets. MATLAB shows better performance in acquiring datasets with higher data contents, whereas C and Python showed a significant drop in the performance, as shown in figure 8*a)*. Also, the MATLAB acquired data with the minimum increase in acquisition time with an increase in the data contents. LabVIEW took the least time to read and write the data files, whereas Python took the most (refer to figures 8 *b) and c)*). Also, the deviation is minimum for LabVIEW and maximum for Python.

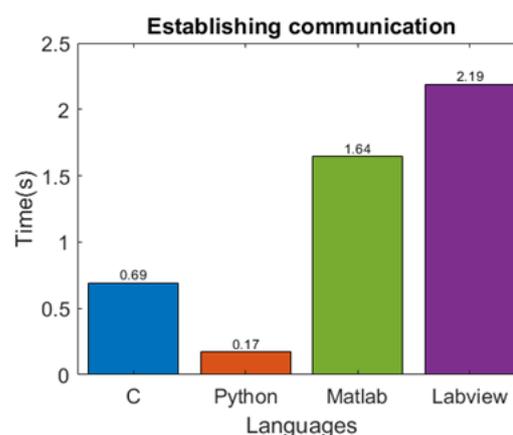

*Figure 7: Time consumed to establish communication to the AWG, DSO, and Arduino.*

2. The runtime performance is measured by recording the time at the several parts of the codes,

   i. Establishing communication to the Instruments: The communication is established to the DSO, wave generator, and microcontroller using the USB port. Python took the least time while LabVIEW took the most, as shown in figure 8.

   ii. Graphics Rendering: The processed data is visualized after each rotation. After each set, graphics are rendered 40 times during the experiment. The data is represented in the form of the boxplot in figure 8*d)*. MATLAB language is performing better, while C has shown poor performance in graphics rendering during the experiment. The C language's poor performance in graphic visualization may be due to interfacing Gnuplot into the C code.

   iii. Complete Experiment: LabVIEW has shown better runtime performance in the experiment without data processing, while MATLAB is performing best in another case(refer to figure 9*a)* and *b)*).

   iv. Completing one set of the process: The timing data is recorded 40 times during a single experiment, representing the time taken to acquire the data 40 times and also process it (only in the case of the experiment when data processing is enabled) in a single set. A total of 1600 datasets are acquired and saved to the PC. The LabVIEW is faster when the experiment is carried out without data processing, as shown in figure 9 *c)*. At the same time, MATLAB performed better when the experiment was carried out with data processing, as shown in figure 9 *d),* respectively. In the second case, C and Python had similar runtime performance, as shown in figure 9 *d)*. However, all the soft tools showed similar standard deviations.

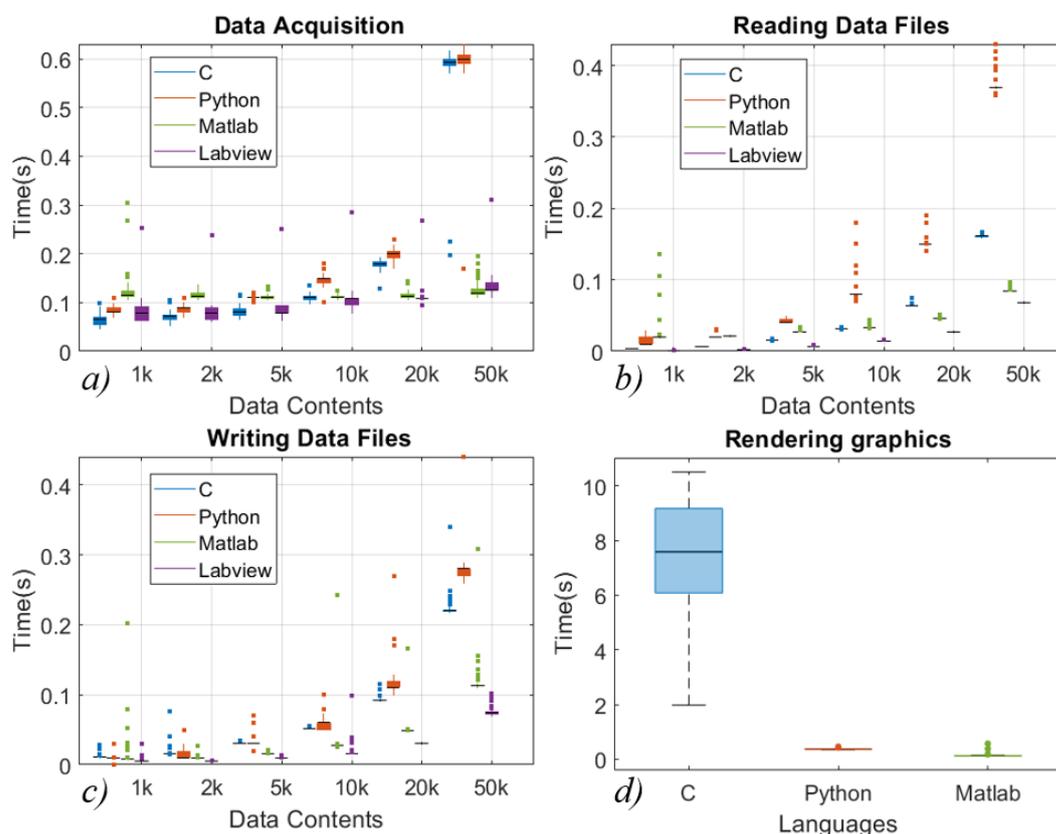

*Figure 8: Runtime performance for a) data acquisition, b) reading data files, c) writing data files, d) rendering the graphics during the experiment.*

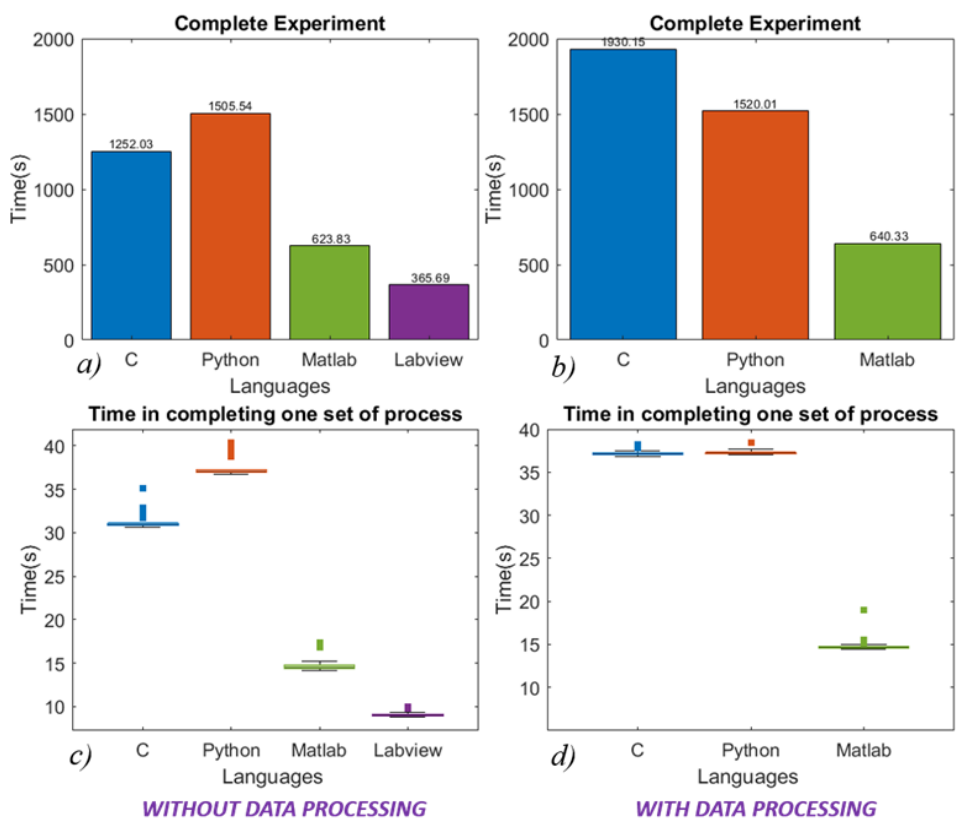

*Figure 9: Runtime performance comparison in a) & b)completing the experiment and c) & d) completing one set of process without data processing(a) & c))and with data processing(b) & d)).*

## Discussion: -

The analysis was performed without optimizing codes for any particular soft tool. The analysis provides a scientific way to analyze the performance of programming language. Based on the analysis, one can choose the best language according to the application requirement. In addition, the library richness, online forums, user experience, and code length play an important role in the selection of a programming language.

Working with C language requires advanced technical knowledge and experience to develop coding skills. The C language provides low-level control in memory management. It can make your applications run faster or become slower over time if one doesn't keep track of memory flow. Running the executable over time with memory leak results in more significant memory leaks that can potentially lead to application crashes. The tracking of performance parameters can provide information about such issues. LabVIEW codes are written in such a way that the code will execute step by step. For the experiment with the data processing, LabVIEW is excluded as it inherently optimizes codes to work in parallel processing. The common mistake made by the researchers while assigning variables is that they don't give a second thought for memory management. It may result in an unoptimized code that will be inefficient in terms of ram utilization, memory utilization and timing performance. It will result in larger inefficiencies while dealing with larger datasets. The major advantages of using Python and MATLAB are that they are supported by a strong academic and research community and online resources.

## Conclusions:-

A UCT system is automated. Four soft tools: C, Python, MATLAB, and LabVIEW, are compared. Apart from the general runtime analysis, two experiments for each language were performed to measure the soft tool's performance in terms of various performance indices. The first experiment was performed without onboard data processing. While the other was performed with data processing.

The IDE for the C language showed a faster launch and response time. The launch performance has significance. If the probability of an operating system or IDE crash (for example, due to some programming fault, high load) is relatively high and the time is limited, C would be preferable. The Python performed better in establishing communication with the instruments. It would be preferable when more instruments or devices are required to be interfaced.

The C language performed better in processor optimization and IO processes. At the same time, Python performed better in the case of RAM usage. The LabVIEW and MATLAB have shown mixed performance. Sometimes, during the experiment, the IO processes/s reduce to a minimum value producing a delay in the experiment. The delay in processes is highest for Python. In the case of Python, for experiment with data processing, the processor utilization and IO processes produce spikes several times that match the number of times graphics are rendered. This suggests that Python executes a higher number of processes for graphics rendering.

In runtime analysis,
i.      In data acquisition, C language is perming better for data contents up to 5k, LabVIEW for contents 10k to 20k, while MATLAB is performing better for higher data contents. However, the performance of all the languages is comparable for the acquisition of datasets having contents up to 10k.
ii.     In reading/writing data files, LabVIEW is performing better. Also, LabVIEW has shown consistent runtime performance. Again, the performance of all the languages is comparable for data contents up to 5k.

iii.     In the complete experiment without data processing, LabVIEW performed better. MATLAB performed better when the experiment was carried out with data processing.

iv.     In graphics rendering, MATLAB is performing better, while Python has shown the least deviation from the mean.

LabVIEW performed better in realtime control of the electromechanical assembly and synchronous data acquisition, and Python performed better runtime efficiency in sensors, and instruments integration, while MATLAB is faster in the simultaneous raw data processing. The C language performed better in optimizing resources, indicating suitability if resources are relatively inferior in processing and storage.

The selection of a programming language primarily depends on the application and convenience. Sometimes, the time runtime efficiency provided by a language can be countered by the time taken to write the codes. The cost may also be a significant factor in selecting a particular language. The MATLAB® and LabVIEW™ are proprietary and closed-source programming environments developed by MathWorks and National Instrument (NI), respectively. A license is required to be purchased, which may be expensive for some users. On the other hand, C and Python are open-source software and are available at no cost.

## Supplementary Data:

A movie (LINK) is provided to show the relative performance of all the IDEs. The codes used in the analysis can be made available on request.

## Acknowledgments:

AK is thankful to the CSIR for the fellowship. MG would like to acknowledge the IMPRINT 2 scheme, Project number: IMP-1348-PHY, Government of India.

## Credit authorship contribution statement

AK: Methodology, Data Acquisition, Investigation, Software, Writing, Visualization, MG: Methodology, Funding, Supervision, Writing.

## Conflicts of Interest: The authors declare no conflict of interest.